\documentclass[runningheads]{llncs}
\usepackage[T1]{fontenc}
\usepackage{graphicx}
\usepackage{booktabs}
\usepackage[misc]{ifsym}
\newcommand{\corr}{(\Letter)}
\usepackage{mwe}

\usepackage{amsmath}
\usepackage{amsthm}
\usepackage{amsfonts}
\usepackage{colortbl}
\usepackage{multirow}
\usepackage{adjustbox}
\usepackage{array}

\usepackage{xcolor}
\usepackage{tcolorbox}

\definecolor{lightyellow}{RGB}{255,255,204}
\definecolor{lightgreen}{RGB}{235,255,235}
\definecolor{lightred}{RGB}{255,204,203}
\definecolor{lightblue}{RGB}{202,225,255} 
\definecolor{validitygreen}{RGB}{0,255,0}

\usepackage{tcolorbox}
\usepackage{listings}
\usepackage{color}
\usepackage{fancyhdr}

\fancypagestyle{firstpagefooter}{
  \fancyhf{}  
  \fancyfoot[C]{\small Accepted for publication at ML4CCE workshop at ECML PKDD 2024.}
}

\definecolor{mygreen}{rgb}{0,0.6,0}
\definecolor{mygray}{rgb}{0.5,0.5,0.5}
\definecolor{mymauve}{rgb}{0.58,0,0.82}

\newcommand{\customsize}{\fontsize{6}{4.2}\selectfont}

\usepackage{fontsize}

\begin{document}

\title{Retrieval-Augmented Instruction Tuning for Automated Process Engineering Calculations : \hspace{5mm} A Tool-Chaining Problem-Solving Framework with Attributable Reflection}

\titlerunning{Reflect, Refine: Tool-Augmented Process Engineering Problem Solving}

\author{Sagar Srinivas Sakhinana\inst{1}\corr \and
Geethan Sannidhi\inst{2}  \and
Venkataramana Runkana\inst{1}}

\authorrunning{Sakhinana, S. S., et al.}
\institute{TCS Research, India \email{\{sagar.sakhinana, venkat.runkana\}@tcs.com}
\and
IIIT Pune, India \email{geethansannidhi20@cse.iiitp.ac.in}}
\tocauthor{Sagar Srinivas Sakhinana, Geethan Sannidhi, Venkataramana Runkana}

\maketitle              

\thispagestyle{firstpagefooter}

\vspace{-9mm}
\begin{abstract} 
The current technology landscape lacks a foundational AI model for solving process engineering calculations. In this work, we introduce a novel autonomous agent framework leveraging Retrieval-Augmented Instruction-Tuning (RAIT) to enhance open, customizable small code language models (SLMs) for these calculations. By combining instruction-tuned code SLMs with Retrieval-Augmented Code Generation (RACG) using external tools, the agent generates, debugs, and optimizes code from natural language specifications. Our approach addresses the limitations of the current lack of a foundational AI model for specialized process engineering tasks and offers benefits of explainability, knowledge-editing, and cost-effectiveness. Additionally, we curate custom datasets of chemical and process engineering problems and solutions to overcome data scarcity. Experimental results show that our framework matches the performance of large-scale proprietary models on benchmark datasets, proving its effectiveness and usability.
\vspace{-3mm}
\vspace{-9mm}
\end{abstract}

\vspace{-4mm}
\section{Introduction} 
\begin{figure*}[!ht]
\centering
\resizebox{0.785\linewidth}{!}{ 
\hspace*{0mm}\includegraphics[keepaspectratio,trim=0.0cm 2.0cm 0cm 0.85cm,clip]{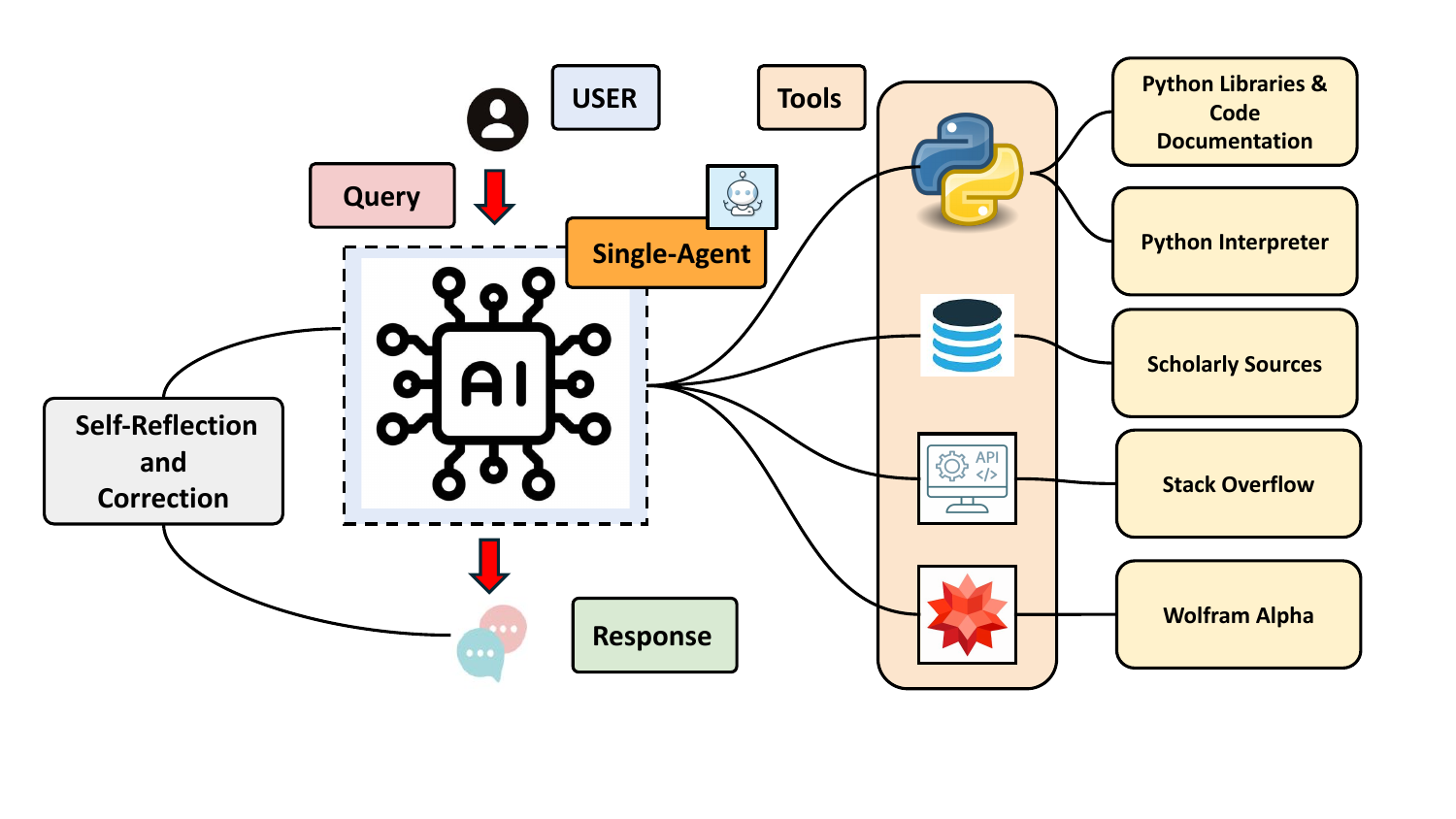} 
}
\vspace{-4mm}
\caption{The figure depicts a single-agent framework for question answering (QA) for solving complex chemical and process engineering calculations.}
\label{fig:figure1}
\vspace{-4mm}
\end{figure*}

\vspace{-4mm}
Basic chemical principles and process calculations, including material and energy balances, heat transfer, phase equilibrium, and reaction kinetics, underpins modeling and optimization in the chemical process industries. These principles are crucial for control and optimization. Process dynamics predict plant behavior and stability, while control ensures desired conditions, quality, and safety. Optimization maximizes profitability, productivity, and efficiency within constraints, contributing to operational excellence. Advances in generative AI, particularly general-purpose large language models (LLMs) like OpenAI's GPT-4 \cite{openai2023gpt4} and Google's Gemini \cite{team2023gemini}, have shown remarkable proficiency in mathematical reasoning and problem-solving skills due to extensive pretraining, revolutionizing various fields. These LLMs can assist engineers in making informed decisions and innovating in the chemical and process industries as decision support tools. Despite these advancements, their ability to solve complex chemical and process calculations remains understudied. Additionally, there is currently no specialized foundational model built for tackling complex chemical and process calculations. There is a need to fine-tune existing pre-trained LLMs on large-scale, domain-specific datasets to develop a new core foundational AI model for solving complex chemical and process calculations. The size and complexity of closed-source LLMs like GPT-4 make it difficult to customize them for specialized tasks and edit knowledge, particularly on consumer hardware with limited computational budgets. In contrast, open small language models (SLMs) like Google Gemma\cite{team2024gemma} and Meta Llama\cite{touvron2023llama} offer domain-specific customization and interpretable analysis. However, SLMs may lack relevant domain-specific mathematical reasoning and problem-solving capabilities compared to proprietary LLMs due to limited relevant pre-trained knowledge. Despite these limitations, a customizable SLM fine-tuned on high-quality domain-specific datasets can achieve performance comparable to large-scale models while offering advantages like explainability and cost-effectiveness. Recently, there has been a surge of interest in exploring tool-augmented SLMs powered by external symbolic tools, numerical software tools, and APIs for program-guided solving that generates executable code to tackle complex mathematical problems. To overcome the drawbacks of limited domain-specific knowledge and better incorporate and utilize external knowledge from databases, we introduce Retrieval-Augmented Instruction-Tuning (RAIT). RAIT technique enhances the capabilities of SLMs by combining Retrieval-Augmented Code Generation (RACG) through external tool usage with instruction-tuning techniques. This approach uses both retrieved information and fine-tuned knowledge for generating code to solve chemical and process engineering problems. However, there are currently no publicly available high-quality labeled datasets of chemical and process engineering problems and solutions to customize SLMs through instruction-tuning. To address this gap, we have curated custom datasets from various scholarly sources suitable for the instruction-tuning of SLMs. The instruction-tuned SLMs utilize both internal parametric pre-trained knowledge and relevant external knowledge from tool invocation (e.g., numerical libraries and code documentation, Stack Overflow, Wolfram Alpha API) with chain-of-thought (CoT)\cite{wei2022chain} reasoning to assist in code generation and debugging. In this work, we introduce an autonomous agent framework that can independently write, debug, and optimize code for solving complex process calculations fundamental to chemical process design, analysis, and optimization. Figure~\ref{fig:figure1} depicts the framework.

\vspace{-4mm}

The agentic framework combines instruction-tuned code-specific language models (such as Google Code Gemma or Meta Code Llama) with Retrieval-Augmented Code Generation (RACG), utilizing the ReAct prompting technique for external knowledge retrieval and reasoning capabilities to generate code from natural language specifications. It follows a five-stage workflow: task planning to analyze the user query and break it into sub-tasks, tool selection to pick the best external tool for each sub-task, parameter extraction from the user query, tool invocation with the extracted parameters, and result integration to synthesize a comprehensive response. The agent can generate executable programs that strategically chain multiple tools to solve complex tasks, eliminating the need for manually designed pipelines. The framework caches successful programs for efficient reuse on similar future tasks. Additionally, it employs an attributable reflection mechanism to handle runtime errors. If such errors occur during program execution, this mechanism identifies the specific tool call causing the error and iteratively revises the program using tool documentation, error information, and program history. This process continues until the program executes successfully or reaches a fixed number of iterations, enabling iterative refinement and improvement of the generated code. The following sections will discuss the framework for solving process engineering problems in great detail, including comprehensive experiments to evaluate the proposed approach. In summary, our contributions are as follows: 

\begin{itemize}
    \vspace{-2mm}
    \item Our work addresses the challenge of applying AI to complex chemical engineering problems by introducing a novel framework. This framework uses a Retrieval-Augmented Instruction-Tuning (RAIT) approach to enhance open SLMs with both domain-specific knowledge from curated datasets and external tools like numerical libraries and APIs. This enables the development of an autonomous agent that can understand natural language instructions, generate, debug, and optimize code to solve intricate process calculations. By combining internal knowledge representation with external tool utilization, this framework aims to provide engineers with a powerful and adaptable tool for process design, analysis, and optimization. We evaluate the effectiveness of our autonomous agent framework in solving complex process calculations through a combination of automatic metrics and a comprehensive user study. Experimental results support our hypothesis that the proposed framework significantly improves solution quality while maintaining high accuracy and efficiency, demonstrating performance comparable to proprietary models.
\end{itemize}

\vspace{-8mm}
\section{Proposed Method} 
\vspace{-3mm}
In this work, we introduce an autonomous agent framework designed to write, debug, and optimize code for systematically solving complex chemical and process calculations, fundamental to the design, analysis, and optimization of the chemical process industry (CPI). The agent performs Retrieval-Augmented Code Generation (RACG), utilizing advanced code-specific small language models (SLMs) such as Google CodeGemma\cite{team2024codegemma} or Meta CodeLlama\cite{codellama} with the ReAct (Reason + Act) prompting\cite{yao2022react} technique to understand and generate code from natural language specifications. Code SLMs lack pre-trained knowledge to generate code for solving chemical and process calculations. Conversely, fine-tuning\cite{zhang2024raft}  by incorporating relevant domain-specific knowledge offers task-specific customization of code SLMs to generate code for solving process calculations. While RACG techniques\cite{lewis2020} allow code SLMs to access external tools\cite{shi2024chain} for code generation, they are often not pre-trained to seamlessly incorporate retrieved information from invoking external tools, leading to less grounded and reliable answers. To address these issues, we utilize the RAIT approach to optimally adapt code SLMs through a fine-tuning technique for domain-specific RACG with tool usage, thus overcoming the limitations of both approaches: limited pre-trained knowledge and the inability to effectively leverage external tools. This leads to improved code generation. The ReAct (Reason + Act) technique enhances the capabilities of code SLMs by prompting them to generate verbal reasoning traces and task-specific actions, helping the code SLM plan and update actions while enabling tool usage to retrieve relevant information. Such tools include APIs like Stack Overflow and Wolfram Alpha, as well as scholarly sources such as code documentation. ReAct employs few-shot learning with examples of task-solving trajectories that include reasoning-action-observation steps, making the agent effective in knowledge-intensive tasks like code generation. This approach allows the agent to reason over external knowledge retrieved from tool usage, extending its capabilities beyond its pre-trained knowledge. We also utilize iterative code generation and refinement through a verify-then-correct approach, involving verification with external critiques to generate feedback and subsequent correction, ensuring the accuracy of the generated code. The autonomous agents operate with a five-stage workflow: task planning, tool selection, parameter extraction from user queries, tool invocation, and result integration. This structured approach maximizes the agents' utility and effectively addresses a wide range of end-user queries through the strategic use of external tools. Task planning involves analyzing the user query to understand the high-level goal and then decomposing it into smaller, manageable sub-tasks. These sub-tasks are organized into a task dependency graph to ensure proper execution order. During tool selection, the most suitable tool is chosen to solve each sub-task from the external tool set. In the parameter extraction stage, the required parameters are extracted from the user query, and in the tool invocation stage, the tool is invoked with these parameters to obtain results. In response generation, the objective is to integrate multiple tool outputs with the code SLM's internal knowledge to generate a comprehensive response. This involves the code language model synthesizing information from the tools and combining it with its fine-tuned knowledge to provide a detailed and accurate response, which may include summarizing the tool outputs. If the initial response is incomplete or requires refinement, an iterative feedback loop\cite{gou2023critic} adjusts the task planning process, ensuring continuous improvement of the agent's performance. For code LMs, tools include external resources such as APIs, software libraries, debugging tools, and knowledge bases like scholarly sources. Leveraging these tools enhances the code generation capabilities of autonomous agents, enabling them to execute, evaluate, and optimize code to meet specific requirements. Additionally, these tools help identify and resolve errors, facilitating better understanding, generation, and manipulation of code for solving complex chemical and process calculations. In summary, the proposed autonomous agent enables code SLMs to act as multi-tool users by generating programs that utilize a chain of tools, aiding in planning and executing complex tasks for solving practical chemical and process calculations. It addresses the limitations of existing ad hoc methods, where tool interaction workflows are manually designed and struggle to generalize to different scenarios. We implement program caching, storing generated programs that successfully complete tasks. This caching allows code SLMs to refer to these programs in future tasks, enabling efficient reuse of previously correct code. This memory of correct code helps reduce redundancy and enhances the efficiency of generating new programs for similar tasks. We illustrate the proposed framework's operation in Tables \ref{table:question}-\ref{table:result_integration} by solving a dynamic modeling problem for a continuous stirred-tank reactor (CSTR) using numerical methods. The task begins with formulating a differential equation for the concentration of a reactant over time. Next, we select the appropriate tools—SciPy for solving differential equations, NumPy for numerical operations, and Matplotlib for plotting results. Parameters are extracted from the user query, and the tools are invoked using a Reason + Act + Observe technique. This approach allows the autonomous agent to effectively integrate the outputs from multiple tools, culminating in a plot of the concentration profile over time. This example shows how the agent generates an accurate response by combining tool outputs with its internal knowledge. 

\vspace{-3mm}
\begin{table*}[ht!]
\begin{tcolorbox}[colback=white, colframe=black, coltitle=black, colbacktitle=white, title=\textbf{Question}.]
\footnotesize
\vspace{-2mm}
A continuous, stirred-tank reactor is initially full of water with the inlet and exit volumetric flow rates of water having the same numerical value. At a particular time, an operator shuts off the water flow and adds caustic solution at the same volumetric flow rate \( q \), but with concentration \( c_i \). If the liquid volume \( V \) is constant, the dynamic model for this process is \( V \frac{dc}{dt} + qc = qc_i \), where \( c(t) \) is the exit concentration. Calculate \( c(t) \) and plot it as a function of time, given \( V = 2 \, \text{m}^3 \), \( q = 0.4 \, \text{m}^3/\text{min} \), and \( c_i = 50 \, \text{kg}/\text{m}^3 \). The initial condition is \( c(0) = 0 \).
\vspace{-2mm}
\end{tcolorbox}
\vspace{-1mm}
\caption{Question describing the dynamic model and task for a continuous, stirred-tank reactor.}
\label{table:question}
\vspace{-6mm}
\end{table*}

\begin{table*}[ht!]
\vspace{-3mm}
\begin{tcolorbox}[colback=white, colframe=black, coltitle=black, colbacktitle=white, title=\textbf{Task Planning}.]
\footnotesize
\vspace{-2mm}
\begin{itemize}
    \item \textbf{Formulate the differential equation}: 
            \vspace{-1mm}
            \begin{itemize}
            \item The high-level goal is to solve the differential equation for a continuous stirred-tank reactor (CSTR) given the problem parameters.
            \vspace{-1mm}
            \item The differential equation is: \( V \frac{dc}{dt} + qc = qc_i \).
        \end{itemize}
    \item \textbf{Convert to standard for}:
        \vspace{-1mm}
        \begin{itemize}
            \item Convert the equation to the standard form for solving: \( \frac{dc}{dt} = \frac{q}{V} (c_i - c) \).
        \end{itemize}
    \item \textbf{Define sub-tasks}:
        \vspace{-1mm}
        \begin{itemize}
            \item  Define the ODE function.
            \vspace{-1mm}
            \item  Set initial conditions.
            \vspace{-1mm}
            \item  Solve the ODE using SciPy.
            \vspace{-1mm}
            \item  Plot the results.
        \end{itemize}
\end{itemize}
\vspace{-3mm}
\end{tcolorbox}
\vspace{-1mm}
\caption{Task planning steps to solve the dynamic model for a stirred-tank reactor.}
\label{table:task_planning}
\vspace{-6mm}
\end{table*}

\begin{table*}[ht!]
\vspace{-3mm}
\begin{tcolorbox}[colback=white, colframe=black, coltitle=black, colbacktitle=white, title=\textbf{Tool Selection}.]
\footnotesize
\vspace{-2mm}
\begin{itemize}
    \item \textbf{SciPys `solve$\_$ivp':}  Selected for solving the initial value problem for the ODE.
    \item \textbf{NumPy:} Used for handling data arrays and calculations.
    \item \textbf{Matplotlib:} Used for plotting the results.
\end{itemize}
\vspace{-4mm}
\end{tcolorbox}
\vspace{-1mm}
\caption{Tool selection for solving the initial value problem for the differential equation of the continuous stirred-tank reactor.}
\label{table:tool_selection}
\vspace{-3mm}
\end{table*}

\begin{table*}[ht!]
\vspace{-3mm}
\begin{tcolorbox}[colback=white, colframe=black, coltitle=black, colbacktitle=white, title=\textbf{Parameter Extraction}.]
\footnotesize
\vspace{-2mm}
\begin{itemize}
    \item \textbf{Volume of the reactor}: \( V = 2 \, \text{m}^3 \).
    \item \textbf{Flow rate}: \( q = 0.4 \, \text{m}^3/\text{min} \).
    \item \textbf{Inlet concentration}: \( c_i = 50 \, \text{kg}/\text{m}^3 \).
    \item \textbf{Initial condition for the concentration}: \( c(0) = 0 \).
\end{itemize}
\vspace{-3mm}
\end{tcolorbox}
\vspace{-1mm}
\caption{Parameter extraction for solving the differential equation of the continuous stirred-tank reactor.}
\label{table:parameter_extraction}
\vspace{-3mm}
\end{table*}

\begin{table*}[ht!]
\vspace{-3mm}
\begin{tcolorbox}[colback=white, colframe=black, coltitle=black, colbacktitle=white, title=\textbf{Tool Invocation with ReAct (Reason + Act) Technique}.]
\footnotesize
\vspace{-2mm}
\begin{itemize}
    \item \textbf{Reason}: 
           \vspace{-1mm}
            \begin{itemize}
            \vspace{-1mm}
            \item  Recognize that the differential equation needs to be solved numerically.
            \vspace{-1mm}
            \item  Understand that `solve$\_$ivp` from SciPy is suitable for this purpose.
        \end{itemize}
        \vspace{-1mm}
    \item \textbf{Act}: 
          \vspace{-1mm}
          \begin{itemize}
          \vspace{-1mm}
           \item  Implement the ODE function.
           \vspace{-1mm}
           \item  Set up the initial conditions and parameters.
           \vspace{-1mm}
           \item  Use `solve$\_$ivp` to solve the ODE.
           \vspace{-1mm}
           \item  Generate a time array for evaluation points.
          \end{itemize} 
          \vspace{-1mm}
    \item \textbf{Reason}:
          \vspace{-1mm}
          \begin{itemize}
           \item Identify the need to visualize the solution..
          \end{itemize}
          \vspace{-1mm}
    \item \textbf{Act}:
          \vspace{-1mm}
          \begin{itemize}
          \vspace{-1mm}
           \item Use Matplotlib to plot the concentration profile over time.
          \end{itemize}
\end{itemize}
\vspace{-5mm}
\end{tcolorbox}
\vspace{-1mm}
\caption{Tool invocation using the ReAct (Reason + Act) technique to solve and visualize the differential equation for the continuous stirred-tank reactor.}
\label{table:tool_invocation}
\vspace{-8mm}
\end{table*}

Note: This illustration is intentionally simplified for ease of understanding. While it demonstrates the key steps of the framework, real-world applications may involve more complex scenarios and additional computational steps. We begin by decomposing a user query \( Q \) into smaller, manageable sub-tasks. Given a natural language query \( Q \), the aim is to enable the code language model to use a sequence of tools from the set \( \mathcal{T} = \{t_1, t_2, \ldots, t_{|\mathcal{T}|}\} \), along with their arguments, to generate an executable program \( C \). The agent initially decides whether it requires tool usage to solve the sub-task.  When tools are not required, the agent relies on its internal pre-trained knowledge to solve the task. If tools are needed, the program chains them together to solve the task. The tool protocols provide meta-information to understand the tools' purpose and usage. The tool protocols \( \mathcal{D} = \{d_1, d_2, \ldots, d_{|\mathcal{D}|}\} \) contain documented protocols \( d_i \) corresponding to each tool \( t_i \in \mathcal{T} \). Each protocol \( d_i \in \mathcal{D} \) provides detailed information about its tool \( t_i \), including argument requirements specifying the inputs needed to run the tool, an overview of the tool's functionality and use cases, and the response schema outlining the expected output structure and type. We retrieve relevant information from external tool usage such as through similarity searches on scholarly articles, web search results, numerical libraries, and Stack Overflow or Wolfram API for mathematical problem-solving tasks. The retrieved information is then used to augment language models for code generation. The detailed tool protocols allow the code language model \( \mathcal{M} \) to learn tool usage, understand the input-output schema, capabilities of various tools, and data flow dependencies, enabling it to generate executable programs \(\mathcal{C}\) that appropriately chain together and utilize multiple tools to solve the natural language task, represented as:

\vspace{-1mm}
\resizebox{0.945\linewidth}{!}{
\begin{minipage}{\linewidth}
\begin{equation}
 \mathcal{C} = \mathcal{M}(Q,  \mathcal{T}, \mathcal{D}, I_c) \nonumber
\end{equation}
\end{minipage}
}

\begin{table*}[ht!]
\vspace{-2mm}
\begin{tcolorbox}[colback=white, colframe=black, coltitle=black, colbacktitle=white, title=\textbf{Result Integration}.]
\footnotesize
\vspace{-2mm}
\begin{verbatim}
import numpy as np
import matplotlib.pyplot as plt
from scipy.integrate import solve_ivp

# Given data
V = 2  # m^3
q = 0.4  # m^3/min
c_i = 50  # kg/m^3

# Define the ODE function
def cstr_ode(t, c, V, q, c_i):
    return (q/V) * (c_i - c)

# initial concentration of A in the reactor
c0 = [0]

# Time span for the solution
t_span = (0, 50) # specifies the start and end times for the integration
t_eval = np.linspace(0, 50, 500) 

# Solve the ODE
solution = solve_ivp(cstr_ode, t_span, c0, args=(V, q, c_i), t_eval=t_eval)

# Extract the results
t = solution.t 
c = solution.y[0] 

# Plot the results
plt.figure(figsize=(10, 6))
plt.plot(t, c, label='c(t)', color='b')
plt.xlabel('Time (minutes)')
plt.ylabel('Concentration of A (kg/m^3)')
plt.title('Concentration of A in the Reactor Over Time')
plt.legend()
plt.grid(True)
plt.show()
\end{verbatim}
\vspace{-5mm}
\end{tcolorbox}
\vspace{-2mm}
\caption{Result integration code to solve and visualize the differential equation for the continuous stirred-tank reactor using Python.}
\label{table:result_integration}
\vspace{-7mm}
\end{table*}

where \( I_c \) indicates a concise instruction prompt provided to the language model for program generation. The generated program \( \mathcal{C} \), by planning a sequence of tool invocations, simplifies the complex task-solving process. By executing the generated program using a code interpreter, the final result (i.e., execution feedback) \( r \) is obtained, which can be represented as:

\vspace{-1mm}
\resizebox{0.945\linewidth}{!}{
\begin{minipage}{\linewidth}
\begin{equation}
r = \text{Execute}(\mathcal{C}) \nonumber
\end{equation}
\end{minipage}
}

\vspace{1mm}
In essence, by leveraging the tool protocols, the code language model can automatically generate programs that strategically chain multi-step tool calls and parse their outputs to solve complex tasks, eliminating the need for manually designed pipelines but potentially encountering runtime errors. If a runtime error occurs during program execution, a reflection mechanism identifies and revises the program to fix the error. If code generation raises an error, the result \( r_j \) obtained during the \( j \)th iteration includes the error message, the faulty code snippet, and the error trace to identify the specific tool causing the error. The code language model localizes the specific tool call \( t_j \) that triggers the error, represented as follows:

\vspace{-4mm}
\resizebox{0.945\linewidth}{!}{
\begin{minipage}{\linewidth}
\begin{equation}
t_j =  \mathcal{M}(Q, \mathcal{T}, I_a, r_j) \nonumber
\end{equation}
\end{minipage}
}

\vspace{1mm}  
where \( I_a \) indicates the instruction prompt to identify and attribute the error to a specific tool. The identified tool \( t_j \) and its documentation \( d_j \), along with the error message, are used to revise the generated program as follows:

\vspace{0mm}
\resizebox{0.945\linewidth}{!}{
\begin{minipage}{\linewidth}
\begin{equation}
\mathcal{C}_j = \mathcal{M}(Q, \mathcal{T}, \mathcal{D}, I_c, r_j, \{(\mathcal{C}_{<j}, r_{<j})\}, d_j) \nonumber
\end{equation}
\end{minipage}
}

\vspace{1mm}  
where \( \mathcal{C}_j \) is the revised program after the \( j \)th iteration of error handling and correction. \( (\mathcal{C}_{<j}, r_{<j}) \) is the history of all previously generated programs and their corresponding results up to, but not including, the \( j \)th iteration. \( I_c \) is the instruction prompt for generating the revised program. This process repeats until the program executes successfully or reaches the predefined maximum number of iterations. In summary, programming with an attributable reflection mechanism\cite{zhang2024evaluating} allows the code LM to introspect on errors, isolate the fault to a specific tool usage, and iteratively refine the program by leveraging tool protocols, error information and the history of prior programs and results. The proposed framework can dynamically generate, execute, and refine programs to solve complex chemical and process industry problems efficiently.

\vspace{-6mm}
\section{Experiments}
\vspace{-4mm}
\paragraph{\textbf{Datasets}:} 
The Retrieval-Augmented Instruction-Tuning (RAIT) technique, which combines tool-augmented RACG and instruction-tuning techniques, can enhance code SLMs' capabilities in handling complex chemical and process engineering problems. To optimize their performance, code SLMs require extensive instruction-tuning using curated datasets to overcome the limitations of their lack of pre-training knowledge, ensuring familiarity with the terminology, methodologies, and typical problem-solving approaches in chemical and process engineering. Solving these problems demands a strong grasp of mathematical concepts and advanced computational tools (e.g., Python libraries). We address this by using instruction-tuning to equip code SLMs with both theoretical knowledge and computational tool learning, allowing them to utilize appropriate tools, thereby augmenting their domain-specific expertise for solving complex chemical and process engineering problems. We have curated a custom instruction-tuning dataset focused on mathematical modeling and numerical algorithms from academic sources and open repositories. The Mathematical and Computational Instruction-Tuning (MathComp) dataset comprises over 7,500 instruction-question-answer triplets designed to adapt code SLMs to computational tool usage, facilitating the generation of executable code to solve ordinary differential equations (ODEs), partial differential equations (PDEs), differential-algebraic equations (DAEs), linear algebra problems, and optimization tasks. Additionally, we have custom-built a Chemical Process Instruction-Tuning (ChemProc) dataset of over 5,000 instruction-QA pairs from scholarly sources and public repositories. This dataset covers topics such as mass and energy balances, thermodynamics, heat and mass transfer, reaction kinetics and reactor design, fluid mechanics and transport phenomena, separation processes, process control, and optimization. The extensive training of code SLMs on these diverse datasets enables them to solve complex chemical and process engineering problems by leveraging domain-specific knowledge and advanced computational techniques. 

\vspace{-3mm}
\paragraph{\textbf{Experimental Settings:}} 
We utilized custom-built MathComp and ChemProc datasets essential for building and evaluating a robust framework capable of handling real-world complex chemical and process engineering problems. These benchmark datasets were split into 70\% training, 15\% validation, and 15\% test sets. Relevant scholarly sources (e.g., textbooks, Python libraries documentation) providing detailed explanations and examples of both numerical methods and chemical and process engineering problems were parsed using a sliding window technique to improve information retrieval. Text chunks were embedded and indexed, with metadata attached to each chunk, including the source document, section title, and chunk position. Images were processed using OpenAI GPT-4 (Omni) to generate text descriptions, which were then indexed for multi-modal search. Scholarly sources play a crucial role in providing high-quality, structured, and comprehensive information that enhances the framework's ability to generate accurate and contextually relevant code solutions for complex engineering problems. We used the open-source BGE embedding method to retrieve relevant passages for knowledge-augmented code generation. To improve retrieval performance for code generation, we fine-tuned an embedding method to rank relevant documents higher by learning semantic relationships. We also used the BGE rerank model to prioritize the most relevant information and fine-tuned for domain adaptation to assign higher relevance scores to the most pertinent passages. The performance of the proposed framework was compared against proprietary models like GPT-4 (Omni), GPT-4 Turbo-preview, Claude-3 Opus, and Google Gemini Pro to ensure a comprehensive evaluation against general-purpose LLMs. We performed instruction-tuning of the code SLMs (Code Gemma-7b-it and Meta Code Llama-7b-hf) using Hugging Face's PEFT library techniques like QLoRA on benchmark datasets for autoregressive code generation on consumer hardware. The code SLMs' fine-tuning leveraged a comprehensive hyperparameter configuration: a batch size of 16, a learning rate of $2 \times 10^{-4}$ adjusted with a constant scheduler over 50 epochs, 100 warmup steps, a weight decay of $1 \times 10^{-4}$, gradient accumulation of 5 steps, and the AdamW optimizer. To ensure efficient parameter updates, we utilized 4-bit QLoRA with a low-rank $r$ of 16, $\alpha$ of 32, and a dropout of 0.05. We utilized NVIDIA GPUs for training, and for evaluation, we performed multiple independent runs and reported ensembled averages.

\vspace{-5mm}
\paragraph{\textbf{Evaluation Metrics:}} 
Code SLMs integrate code generation skills, such as the ability to generate and refine code, within the broader framework of tool learning. This enables them to better handle complex mathematical reasoning and programming tasks, thereby improving problem-solving skills in chemical and process calculations. Our study employs various evaluation metrics to assess the effectiveness of the code SLMs' tool learning capabilities\cite{qu2024tool} across different stages: task planning, tool selection, tool calling, and response generation. For task planning, we use Tool Usage Awareness, Pass Rate, and Accuracy. Tool Usage Awareness measures the ability to identify if a query requires an external tool, expressed as \( \text{Awareness} = \frac{\text{Number of Correct Identifications}}{\text{Total Number of Queries}} \). Pass Rate assesses task planning effectiveness, calculated by \( \text{Pass Rate} = \frac{\text{Number of Successfully Completed Tasks}}{\text{Total Number of Tasks}} \). Accuracy evaluates the precision of the plan, calculated as \( \text{Accuracy} = \frac{\text{Number of Correct Plans}}{\text{Total Number of Plans}} \). For tool selection, we use Recall@K, NDCG@K, and COMP@K. Recall@K measures the proportion of selected top-K tools present in the set of ground-truth tools, formulated as \( \text{Recall@K} = \frac{1}{|Q|} \sum_{q=1}^{|Q|} \frac{|T^K_q \cap T^*_q|}{|T^*_q|} \), where \( |Q| \) is the set of queries, \( T^*_q \) is the set of relevant tools for the query \( q \), and \( T^K_q \) is the top-K tools for the query \( q \) selected by the framework. NDCG@K considers the proportion and positions of positive tools, with Discounted Cumulative Gain (DCG@K) calculated as \( \text{DCG}_q@K = \sum_{i=1}^K \frac{2^{g_i} - 1}{\log_2 (i+1)} \), where \( g_i \) is the graded relevance score (assigned by human evaluators) at position \( i \le K\) in the top-K list. IDCG@K is Ideal Discounted Cumulative Gain. COMP@K assesses whether the top-K selected tools form a complete set with respect to the ground-truth set, defined as \( \text{COMP@K} = \frac{1}{|Q|} \sum_{q=1}^{|Q|} I(\Phi_q \subseteq \Psi^K_q) \). \(\Phi_q\) is the ground-truth tool set for query \(q\), and \(\Psi^K_q\) is the top-K tools retrieved for query \(q\). For tool calling, we use Consistency with Stipulations, Correctness of Parameter Extraction, and Error Handling as evaluation metrics. Consistency with Stipulations measures how well the provided parameters match the tool's documentation requirements, calculated as \( \left( \frac{\text{Number of parameters consistent with the stipulations}}{\text{Total number of parameters required}} \right) \). Correctness of Parameter Extraction evaluates the accuracy in extracting correct parameters from the user query, defined as \( \left( \frac{\text{Number of correctly extracted parameters}}{\text{Total number of parameters}} \right) \). Error Handling assesses the system's ability to manage errors during tool calling, measured as \( \left( \frac{\text{Number of errors handled successfully}}{\text{Total number of errors encountered}} \right) \). For response generation, we use BLEU, ROUGE-L, and Exact Match. BLEU is calculated as \( \text{BLEU} = BP \cdot \exp \left( \sum_{n=1}^{N} w_n \log p_n \right) \). ROUGE-L is calculated as \( \text{ROUGE-L} = F_{\beta} = \frac{(1 + \beta^2) \cdot \text{LCS-precision} \cdot \text{LCS-recall}}{\text{LCS-precision} + \beta^2 \cdot \text{LCS-recall}} \), where LCS-Precision is the ratio of the length of the Longest Common Subsequence (LCS) to the total number of words in the candidate response and LCS-Recall is the ratio of the length of the LCS to the total number of words in the reference response. Exact Match is calculated as \( \text{Exact Match} = \frac{\text{Number of Exact Matches}}{\text{Total Number of Responses}} \). These evaluation metrics provide a comprehensive framework for evaluating the performance of code SLMs in tool learning tasks, ensuring that they can effectively plan, select, call, and integrate tools to enhance their problem-solving capabilities.

\vspace{-3mm}
\paragraph{\textbf{Experimental Results:}} 
The table \ref{tab:task_planning} compares algorithms in task planning by Tool Usage Awareness (TUA), Pass Rate (PR), and Accuracy (Acc), all in \%. TUA ranges from 0\% (failure) to 100\% (perfect identification). PR ranges from 0\% (none correct) to 100\% (all correct). Acc ranges from 0\% (none identified correctly) to 100\% (all identified correctly). The table \ref{tab:tool_selection} compares algorithms in tool selection by Recall, NDCG, and COMP metrics. Recall ranges from 0 (none relevant) to 100 (all relevant). NDCG ranges from 0 (worst) to 1 (ideal). COMP ranges from 0 (none selected) to 100 (all selected).

\begin{table}[h!]
\vspace{-3mm}
\centering
\renewcommand{\arraystretch}{1.0}
\resizebox{0.965\textwidth}{!}{
\begin{tabular}{lccc}
\toprule
\textbf{Algorithm} & \textbf{Tool Usage Awareness (\%)} & \textbf{Pass Rate (\%)} & \textbf{Accuracy (\%)} \\
\midrule
GPT-4 (Omni) & 95.12 & 90.45 & 92.78 \\ 
GPT-4 Turbo-preview & 93.34 & 88.56 & 90.23 \\ 
Claude-3 Opus & 92.47 & 87.89 & 89.65 \\ 
Google Gemini Pro & 94.15 & 89.37 & 91.58 \\ \hline
Proposed Framework & 86.89 & 82.34 & 84.21 \\ 
\bottomrule
\end{tabular}
}
\vspace{0mm}
\caption{The table provides a comprehensive overview of performance evaluation metrics and their corresponding results for LLMs in the context of task planning.}
\label{tab:task_planning}
\vspace{-10mm}
\end{table}

\begin{table}[h!]
\centering
\renewcommand{\arraystretch}{1.0}
\resizebox{0.65\textwidth}{!}{
\begin{tabular}{lccc}
\toprule
\textbf{Algorithm} & \textbf{Recall (\%)} & \textbf{NDCG} & \textbf{COMP (\%)} \\
\midrule
GPT-4 (Omni) & 92.54 & 0.88 & 90.12 \\ 
GPT-4 Turbo-preview & 90.23 & 0.85 & 88.45 \\ 
Claude-3 Opus & 89.67 & 0.84 & 87.34 \\ 
Google Gemini Pro & 91.12 & 0.86 & 89.76 \\ \hline
Proposed Framework & 83.45 & 0.75 & 81.23 \\ 
\bottomrule
\end{tabular}
}
\vspace{1mm}
\caption{The table offers a summary of key performance evaluation metrics and corresponding results for LLMs performance in tool selection}
\label{tab:tool_selection}
\vspace{-10mm}
\end{table}

\begin{table}[h!]
\centering
\renewcommand{\arraystretch}{1.0}
\resizebox{0.575\textwidth}{!}{
\begin{tabular}{lccc}
\toprule
\textbf{Algorithm} & \textbf{Cons (\%)} & \textbf{PE (\%)} & \textbf{EH (\%)} \\
\midrule
GPT-4 (Omni) & 93.12 & 91.34 & 90.67 \\ 
GPT-4 Turbo-preview & 91.45 & 89.78 & 88.56 \\ 
Claude-3 Opus & 90.23 & 88.45 & 87.34 \\ 
Google Gemini 1.0 Pro & 92.56 & 90.12 & 89.89 \\ \hline
Proposed Framework & 84.23 & 82.45 & 81.67 \\ 
\bottomrule
\end{tabular}
}
\vspace{1mm}
\caption{The table presents LLM performance in tool calling, evaluated using key metrics.}
\label{tab:tool_calling}
\vspace{-10mm}
\end{table}

\begin{table}[h!]
\centering
\renewcommand{\arraystretch}{1.0}
\resizebox{0.705\textwidth}{!}{
\begin{tabular}{lccc}
\toprule
\textbf{Algorithm} & \textbf{BLEU} & \textbf{ROUGE-L} & \textbf{Exact Match (\%)} \\
\midrule
GPT-4 (Omni) & 0.87 & 0.85 & 90.37 \\
GPT-4 Turbo-preview & 0.85 & 0.83 & 88.49 \\
Claude-3 Opus & 0.84 & 0.82 & 87.03 \\
Google Gemini 1.0 Pro & 0.86 & 0.84 & 87.57 \\ \hline
Proposed Framework & 0.78 & 0.76 & 80.92 \\
\bottomrule
\end{tabular}
}
\vspace{1mm}
\caption{The table summarizes performance metrics for LLMs in response generation.}
\label{tab:response_generation}
\vspace{-7mm}
\end{table}

The table \ref{tab:tool_calling} compares algorithms in tool calling by Consistency with Stipulations (Cons), Correctness of Parameter Extraction (PE), and Error Handling (EH), all in \%. Cons ranges from 0\% (none meet requirements) to 100\% (all meet requirements). PE ranges from 0\% (none correct) to 100\% (all correct). EH ranges from 0\% (ineffective) to 100\% (effective). The table \ref{tab:response_generation} compares algorithms in response generation by BLEU, ROUGE-L, and Exact Match (EM). BLEU ranges from 0 (worst) to 1 (perfect). ROUGE-L ranges from 0 (no common subsequence) to 1 (perfect common subsequence). EM ranges from 0\% (none match) to 100\% (all match). The experimental results indicate that the proposed framework is effective; however, it lags slightly behind the proprietary models. We conducted several ablation studies to thoroughly evaluate our framework, which leverages the RAIT technique to enhance code SLMs for solving complex chemical and process engineering calculations. The ablation studies aim to isolate and quantify the contributions of key components: the ReAct (Reason + Act) prompting technique (without ReAct), integration of external knowledge (without Ext. Know), fine-tuning of code SLMs (without SLM Tuning), iterative code refinement (without Iter Refine) process with the attributable reflection mechanism, and program caching (without Caching). By systematically removing or disabling each component and observing the resulting performance changes, we measure their impact on key metrics. The baseline is the complete RAIT framework with all components enabled. These ablation studies ensure a thorough understanding of the framework's strengths, demonstrating how each component contributes to the overall performance in solving specialized engineering problems. The ablation studies presented in Tables \ref{tab:ablation_task_planning}, \ref{tab:ablation_tool_selection}, \ref{tab:ablation_tool_calling}, and \ref{tab:ablation_response_generation} provide a comprehensive evaluation of the proposed RAIT framework for enhancing code SLMs in solving complex chemical and process engineering calculations.

\begin{table}[h!]
\vspace{-3mm}
\centering
\renewcommand{\arraystretch}{1.0}
\resizebox{0.935\textwidth}{!}{
\begin{tabular}{lccc}
\toprule
\textbf{Algorithm} & \textbf{Tool Usage Awareness (\%)} & \textbf{Pass Rate (\%)} & \textbf{Accuracy (\%)} \\
\midrule
w/o ReAct & 67.36 & 63.71 & 66.51 \\ 
w/o Ext. Know & 64.51 & 61.72 & 56.12 \\ 
w/o Iter Refine & 76.89 & 70.64 & 73.74 \\ 
w/o SLM Tuning & 37.72 & 33.11 & 34.28 \\  
w/o Caching & 83.10 & 78.81 & 79.49 \\ \hline
Proposed Framework & 86.89 & 82.34 & 84.21 \\  
\bottomrule
\end{tabular}
}
\vspace{1mm}
\caption{The table highlights the performance of ablated variants compared to the baseline in task planning.}
\label{tab:ablation_task_planning}
\vspace{-11mm}
\end{table}

\begin{table}[h!]
\centering
\renewcommand{\arraystretch}{1.0}
\resizebox{0.615\textwidth}{!}{
\begin{tabular}{lccc}
\toprule
\textbf{Algorithm} & \textbf{Recall (\%)} & \textbf{NDCG} & \textbf{COMP (\%)} \\
\midrule
w/o ReAct & 65.09 & 0.58 & 61.73 \\
w/o Ext. Know & 58.41 & 0.51 & 53.61 \\
w/o Iter Refine & 72.60 & 0.66 & 69.70 \\
w/o SLM Tuning & 37.73 & 0.30 & 35.67 \\
w/o Caching & 79.69 & 0.71 & 77.09 \\ \hline
Proposed Framework & 83.45 & 0.75 & 81.23 \\ 
\bottomrule
\end{tabular}
}
\vspace{1mm}
\caption{The table shows the performance of ablated variants compared to the baseline in tool selection.}
\label{tab:ablation_tool_selection}
\vspace{-12mm}
\end{table}

\begin{table}[h!]
\centering
\renewcommand{\arraystretch}{1.0}
\resizebox{0.545\textwidth}{!}{
\begin{tabular}{lccc}
\toprule
\textbf{Algorithm} & \textbf{Cons (\%)} & \textbf{PE (\%)} & \textbf{EH (\%)} \\
\midrule
w/o ReAct & 64.53 & 64.58 & 61.35 \\
w/o Ext. Know & 58.52 & 55.17 & 58.01 \\
w/o Iter Refine & 73.53 & 71.29 & 69.57 \\
w/o SLM Tuning & 36.91 & 29.48 & 31.90 \\
w/o Caching & 80.24 & 78.87 & 77.65 \\ \hline
Proposed Framework & 84.23 & 82.45 & 81.67 \\ 
\bottomrule
\end{tabular}
}
\vspace{1mm}
\caption{The table outlines the performance of ablated variants compared to the baseline in tool calling.}
\label{tab:ablation_tool_calling}
\vspace{-14mm}
\end{table}

\clearpage
\newpage

\begin{table}[h!]
\centering
\renewcommand{\arraystretch}{1.0}
\resizebox{0.685\textwidth}{!}{
\begin{tabular}{lccc}
\toprule
\textbf{Algorithm} & \textbf{BLEU} & \textbf{ROUGE-L} & \textbf{Exact Match (\%)} \\
\midrule
w/o Caching & 0.75 & 0.72 & 76.58 \\
w/o Iter Refine & 0.66 & 0.65 & 71.88 \\
w/o ReAct & 0.59 & 0.58 & 64.58 \\
w/o SLM Tuning & 0.37 & 0.405 & 35.21 \\
w/o Ext. Know & 0.56 & 0.56 & 52.78 \\ \hline
Proposed Framework & 0.78 & 0.76 & 80.92 \\ 
\bottomrule
\end{tabular}
}
\caption{The table presents the performance of ablated variants compared to the baseline in response generation.}
\label{tab:ablation_response_generation}
\vspace{-6mm}
\end{table}

The ablation study experimental results clearly demonstrate that the complete RAIT framework (baseline) outperforms the ablated variants across various metrics. This highlights the synergistic effect of the framework's components and underscores the importance of incorporating all aspects for optimal performance in specialized engineering tasks.
A comprehensive human evaluation\cite{xiao2024human} approach is developed to evaluate the proposed RAIT framework. This framework is designed to augment code SLMs' capabilities in solving complex chemical and process engineering calculations. The evaluation involves criteria including accuracy, relevance, coherence, comprehensiveness, and usability, assessed by both domain experts and general technical evaluators. Accuracy measures the correctness of the solutions, relevance assesses the appropriateness of the responses and selected tools, coherence evaluates the logical flow and readability, comprehensiveness checks the coverage and detail level, and usability determines the practicality and ease of integration of the generated code into workflows. Detailed guidelines ensure consistent scoring across these criteria, using a 1 (Poor) to 5 (Excellent) scale for each criterion. Both quantitative scores and qualitative feedback provide in-depth insights into the framework's performance. The process involves expert evaluators analyzing results to identify strengths and weaknesses based on benchmark datasets. The table \ref{tab:overall_performance_metrics} presents the overall performance metrics for the RAIT framework, showing the mean scores across five evaluation criteria: accuracy, relevance, coherence, comprehensiveness, and usability. The scores range from 4.00 to 4.22 on a scale of 1 (Poor) to 5 (Excellent).

\vspace{-1mm}
\begin{table}[h!]
\centering
\begin{tabular}{|l|c|}
\hline
\textbf{Criterion} & \textbf{Mean Score} \\
\hline
Accuracy           & 4.22 \\
Relevance          & 4.00 \\
Coherence          & 4.22 \\
Comprehensiveness  & 4.00 \\
Usability          & 4.11 \\
\hline
\end{tabular}
\vspace{1mm}
\caption{The table shows human-evaluation performance metrics for RAIT framework.}
\label{tab:overall_performance_metrics}
\vspace{-7mm}
\end{table}

We demonstrate the practical application and effectiveness of the RAIT framework in solving a variety of chemical engineering calculations. Table \ref{tab:gas_law} uses the ideal gas law to calculate $\text{CO}_{2}$ volume. Table \ref{tab:biosynthesis} demonstrates modeling biosynthesis: $\text{CO}_{2}$ utilization and product formation. Table \ref{tab:humidity} presents the solution for psychrometric calculations in industrial dehumidification. Table \ref{tab:evaporator} analyzes a single-effect evaporator, calculating heat load and steam economy.

\begin{table*}[ht!]
\vspace{-4mm}
\begin{tcolorbox}[colback=white, colframe=black, coltitle=black, colbacktitle=white, title=\textbf{\footnotesize Calculate the volume occupied by 88 lb of CO$_{2}$ at 15$^{\circ}$C and a pressure of 32.2 ft of water}.]
\footnotesize
\vspace{-3mm}
To calculate the volume of 88 lb of CO$_{2}$ at 15$^{\circ}$C and a pressure of 32.2 ft of water, use the Ideal Gas Law: \( PV = nRT \). Convert units: 15$^{\circ}$C to 288.15 K and 32.2 ft of water to 13.94 psi using \( 1 \, \text{ft of water} = 0.433 \, \text{psi} \). Calculate moles of CO$_{2}$ using its molar mass (44.01 g/mol) and \( 1 \, \text{lb} = 453.592 \, \text{g} \): \( 88 \, \text{lb} \times 453.592 \, \text{g/lb} / 44.01 \, \text{g/mol} = 907.18 \, \text{moles} \). Convert pressure to atmospheres: \( 13.94 \, \text{psi} = 0.948 \, \text{atm} \). Use the ideal gas constant \( R = 0.0821 \, \text{L} \cdot \text{atm/mol} \cdot \text{K} \) and convert the final volume from liters to cubic feet using \( 1 \, \text{ft}^{3} = 28.317 \, \text{L} \). Solve for the volume: \( V = \frac{nRT}{P} = 21,668 \, \text{L} \) and convert to cubic feet: \( 21,668 \, \text{L} \times \left(\frac{1 \, \text{ft}^{3}}{28.317 \, \text{L}}\right) = 765.8 \, \text{ft}^{3} \). Thus, the volume is approximately 765.8 $\text{feet}^{3}$.
\vspace{-3mm}
\end{tcolorbox}
\vspace{0mm}
\caption{The framework calculates $CO_{2}$ volume using the Ideal Gas Law, including unit conversions and step-by-step problem-solving approach.}
\label{tab:gas_law}
\vspace{-5mm}
\end{table*}

\begin{table*}[ht!]
\vspace{-5mm}
\begin{tcolorbox}[colback=white, colframe=black, coltitle=black, colbacktitle=white, title=\textbf{\footnotesize NADH (nicotinamide adenine dinucleotide) supplies hydrogen in living cells for biosynthesis reactions such as \( \text{CO}_2 + 4\text{H} \rightarrow \text{CH}_2\text{O} + \text{H}_2\text{O} \). If you saturate 1 L of deaerated water with \( \text{CO}_2 \) gas at \( 20^\circ \text{C} \) (the solubility is 1.81 g \( \text{CO}_2 \)/L) and add enough NADH to provide 0.057 g of H into a bioreactor used to imitate the reactions in cells, and obtain 0.7 g of \( \text{CH}_2\text{O} \), what is the extent of reaction for this reaction? Use the extent of reaction to determine the number of grams of \( \text{CO}_2 \) left in solution.}.]
\footnotesize
\vspace{-2mm}
To calculate the extent of reaction, we use the amount of \( \text{CH}_2\text{O} \) produced. First, we convert the mass of \( \text{CH}_2\text{O} \) to moles: \( 0.7 \text{ g} / 30.03 \text{ g/mol} = 0.0233 \text{ mol} \). The stoichiometric coefficient for \( \text{CH}_2\text{O} \) is 1, so the extent of reaction \( \xi \) is 0.0233 mol. The initial moles of \( \text{CO}_2 \) in 1 L of water are \( 1.81 \text{ g} / 44.01 \text{ g/mol} = 0.0411 \text{ mol} \). Using the extent of reaction, the moles of \( \text{CO}_2 \) left in solution are \( 0.0411 - 0.0233 = 0.0178 \text{ mol} \). Finally, converting back to grams: \( 0.0178 \text{ mol} \times 44.01 \text{ g/mol} = 0.784 \text{ g} \). Thus, the number of grams of \( \text{CO}_2 \) left in solution is 0.784 g.
\vspace{-2mm}
\end{tcolorbox}
\vspace{0mm}
\caption{The framework calculates the reaction extent and the remaining $\text{CO}_{2}$ in a cellular biosynthesis process (stoichiometric analysis of $\text{CO}_{2}$ to $\text{CH}_2\text{O}$ conversion).}
\label{tab:biosynthesis}
\vspace{-6mm}
\end{table*}
 
\begin{table*}[ht!]
\vspace{-4mm}
\begin{tcolorbox}[colback=white, colframe=black, coltitle=black, colbacktitle=white, title=\textbf{\footnotesize A dryer must remove 200 kg of H$_2$O per hour from a certain material. Air at 22°C and 50\% relative humidity enters the dryer and leaves at 72°C and 80\% relative humidity. What is the weight (in kilograms) of bone-dry air used per hour? The barometer reads 103.0 kPa.}.]
\footnotesize
\vspace{-2mm}
We first calculate the humidity ratios(represents the mass of water vapor per unit mass of dry air) at the inlet and outlet conditions using the formula $W = \frac{0.622 \times P_v}{P - P_v}$ where $P_v$ is the partial pressure of water vapor, $P$ is  is the total pressure and 0.622 is the ratio of the molecular weight of water vapor to the molecular weight of dry air . For the inlet conditions (22°C, 50\% relative humidity), the saturation pressure is approximately 2.64 kPa, giving $P_{v,\text{inlet}} = 1.32$ kPa and $W_{\text{inlet}} \approx 0.00806$ kg H$_2$O/kg dry air. For the outlet conditions (72°C, 80\% relative humidity), the saturation pressure is approximately 31.84 kPa, giving $P_{v,\text{outlet}} = 25.47$ kPa and $W_{\text{outlet}} \approx 0.204$ kg H$_2$O/kg dry air. Using the mass balance equation 
$\dot{m}_{H_2O} = \dot{m}_{\text{dry air}} \times (W_{\text{outlet}} - W_{\text{inlet}})$ and solving for $\dot{m}_{\text{dry air}}$, we find 
$ \dot{m}_{\text{dry air}} = \frac{200}{0.204 - 0.00806} \approx 1021.33 \, \text{kg/hr}$. $W_{\text{outlet}} - W_{\text{inlet}}$ represents the increase in water vapor content per unit mass of dry air as the air passes through the dryer and $\dot{m}_{\text{dry air}}$ is the mass flow rate of the dry air that is used in the drying process. Thus, the weight of bone-dry air used per hour is approximately 1021.33 kg.
\vspace{-2mm}
\end{tcolorbox}
\vspace{0mm}
\caption{The framework calculates the dry air mass flow in an industrial drying process (air consumption in a controlled humidity drying environment).}
\label{tab:humidity}
\vspace{-10mm}
\end{table*}

\clearpage
\newpage

\begin{table*}[ht!]
\vspace{-4mm}
\begin{tcolorbox}[colback=white, colframe=black, coltitle=black, colbacktitle=white, title=\textbf{\footnotesize A single effect evaporator is to concentrate 20000 kg/h of a solution having a concentration of 5\% salt to a concentration of 20\% salt by weight. Steam is fed to the evaporator at a pressure corresponding to the saturation temperature of 399 K (126$^\circ$C). The evaporator is operating at atmospheric pressure, and the boiling point rise is 7 K. Calculate the heat load and steam economy.)}]
\footnotesize
\vspace{-2mm}
Given data includes the initial flow rate of the solution at 20000 kg/h, an initial concentration of 5\%, and a final concentration of 20\%. The steam temperature is 399 K and the boiling point of the solution is 373 K with a boiling point rise of 7 K. First, we calculate the amount of water to be evaporated using the mass balance for salt: $F \cdot C_f = (F - V) \cdot C_p$, rearranging for $V$ gives us $V = \frac{20000 \cdot (0.05 - 0.20)}{0.20} = 15000$ kg/h. Next, we calculate the heat load, using the latent heat of vaporization of water (approximately 2260 kJ/kg): $Q = V \cdot \lambda = 15000 \cdot 2260 = 33900000$ kJ/h or 33900 kW. Then, we determine the steam economy, defined as the amount of water evaporated per unit mass of steam used. Assuming the steam condenses at its saturation temperature (126$^\circ$C or 399 K) with the latent heat of steam also approximately 2260 kJ/kg, the amount of steam used is $S = \frac{Q}{\lambda_s} = \frac{33900000}{2260} = 15000$ kg/h. The steam economy is then calculated as $SE = \frac{V}{S} = \frac{15000}{15000} = 1.0$. In summary, the heat load is 33900 kW (or 33900000 kJ/h) and the steam economy is 1.0.
\vspace{-2mm}
\end{tcolorbox}
\vspace{-2mm}
\caption{The framework determines the heat load and steam economy for an atmospheric pressure evaporator.}
\label{tab:evaporator}
\vspace{-8mm}
\end{table*}

We introduce `Needle in a Haystack' experiments to evaluate the RAIT framework using 100 Q\&A pairs from benchmark test sets. These experiments assess the framework's ability to extract specific, relevant information from large knowledge bases to solve engineering problems. Key performance metrics include precision (the fraction of retrieved documents that are relevant to the query), emphasizing minimization of false positives, and recall (the fraction of all relevant documents that were successfully retrieved), emphasizing minimization of false negatives. Additional metrics are the F1 score (harmonic mean of precision and recall), accuracy of generated answers, relevance of retrieved documents, and coherence of responses. Results in Table \ref{tab:needle_haystack_results} show that the RAIT framework achieves performance comparable to larger models in specialized tasks, demonstrating its effectiveness in complex information retrieval and problem-solving scenarios.

\begin{table}[h!]
\vspace{-2mm}
\centering
\renewcommand{\arraystretch}{1.0}
\resizebox{0.85\textwidth}{!}{
\begin{tabular}{|l|c|c|c|c|c|}
\hline
\textbf{Metric} & \textbf{GPT-4(O)} & \textbf{GPT-4(T)} & \textbf{Claude Opus} & \textbf{Google Gemini} & \textbf{RAIT} \\
\hline
\textbf{Precision} & 85\% & 83\% & 82\% & 84\% & 77.45\% \\
\hline
\textbf{Recall} & 88\% & 86\% & 85\% & 87\% &  78.04\% \\
\hline
\textbf{F1 Score} & 86\% & 84\% & 83\% & 85\% & 79.07\% \\
\hline
\textbf{Accuracy} & 91\% & 89\% & 88\% & 90\% & 83.31\% \\
\hline
\textbf{Relevance} & 87\% & 85\% & 84\% & 86\% & 79.46\% \\
\hline
\textbf{Coherence} & 90\% & 88\% & 87\% & 89\% & 78.80\% \\
\hline
\end{tabular}
}
\vspace{1mm}
\caption{The table shows results of the ``Needle in a Haystack" experiments.}
\label{tab:needle_haystack_results}
\vspace{-13mm}
\end{table}

\section{Conclusion}
\vspace{-4mm}
In conclusion, the proposed RAIT technique effectively enhances code SLMs for complex chemical engineering calculations. By combining retrieval-augmented generation with instruction-tuning, the RAIT technique addresses the limitations of code SLMs in specialized tasks. Experiments demonstrate the effectiveness of RAIT's techniques, rivaling larger language models in domain-specific applications. Our work offers a promising direction for developing efficient, specialized AI tools within the chemical process industry. Future work should focus on developing more comprehensive and diverse datasets for instruction-tuning, incorporating a wider range of external tools and APIs, and exploring advanced techniques for knowledge retrieval and integration. Additionally, investigating ways to improve the framework's explainability and interpretability would be valuable for building trust and facilitating adoption in industrial settings.

\vspace{-4mm}
\bibliographystyle{splncs04}
\bibliography{mybibliography}

\end{document}